\documentclass[camera,letterpaper,nomarginnotes,nonarrowgutter,hyphens]{jpaper}

\usepackage[colorlinks]{hyperref}
\usepackage[dvipsnames]{xcolor}
\usepackage{soul}
\usepackage{xspace}
\usepackage{multirow}
\usepackage{tabularx}
\usepackage{tikz}
\usepackage[bottom]{footmisc}

\usepackage[most]{tcolorbox}
\usepackage{amsmath}
\usepackage{enumitem}

\usepackage{subcaption}
\usepackage{fancyhdr}
\usepackage{mathtools}

\usepackage{booktabs}

\usepackage{soul}
\usepackage{stfloats}
\definecolor{darkspringgreen}{rgb}{0.09, 0.45, 0.27}
\definecolor{denim}{rgb}{0.08, 0.38, 0.74}
\definecolor{darkolivegreen}{rgb}{0.33, 0.42, 0.18}
\definecolor{tangerine}{rgb}{0.95, 0.52, 0.0}
\definecolor{mahogany}{rgb}{0.75, 0.25, 0.0}
\definecolor{coolblack}{rgb}{0.0, 0.18, 0.39}
\definecolor{darkpink}{rgb}{0.88, 0.28, 0.54}

\definecolor{seagreen}{rgb}{0.18, 0.55, 0.34}

\usepackage[colorinlistoftodos,prependcaption,textsize=scriptsize,textwidth=13mm]{todonotes} %
\usepackage{marginnote} 
\let\marginpar\marginnote
\newcommand\revmid[1]{\todo[linecolor=magenta,backgroundcolor=magenta!15,bordercolor=magenta]{\textcolor{black}{\textbf{#1}}}}

\renewcommand\revmid[1]{}

\definecolor{pred}{rgb}{0.7843, 0.0039, 0.3137} %

\definecolor{darkpink}{rgb}{0.88, 0.28, 0.54}
\definecolor{forestgreen}{rgb}{0.0, 0.27, 0.13}
\definecolor{amber}{rgb}{1.0, 0.49, 0.0}

\newcommand\proposal{PAPI\xspace}

\newcolumntype{Y}{>{\centering\arraybackslash}X}
\usepackage{tikz}

\newcommand{\squishlist}{
 \begin{list}{$\circ$}
  { \setlength{\itemsep}{0pt}
     \setlength{\parsep}{0pt}
     \setlength{\topsep}{3pt}
     \setlength{\partopsep}{0pt}
     \setlength{\leftmargin}{1em}
     \setlength{\labelwidth}{1em}
     \setlength{\labelsep}{0.5em} } }

\newcommand{\squishend}{
  \end{list}  }

\usepackage[compact]{titlesec}
\titlespacing\section{0pt}{4pt plus 2pt minus 2pt}{4pt plus 2pt minus 2pt}
\titlespacing\subsection{0pt}{4pt plus 2pt minus 2pt}{4pt plus 2pt minus 2pt}
\titlespacing\subsubsection{0pt}{4pt plus 2pt minus 2pt}{4pt plus 2pt minus 2pt}
\makeatletter
\g@addto@macro{\normalsize}{%
  \setlength{\abovedisplayskip}{5pt plus 0.5pt minus 1pt}
  \setlength{\belowdisplayskip}{5pt plus 0.5pt minus 1pt}
  \setlength{\abovedisplayshortskip}{3pt}
  \setlength{\belowdisplayshortskip}{3pt}
  \setlength{\intextsep}{3pt plus 1pt minus 1pt}
  \setlength{\textfloatsep}{3pt plus 1pt minus 1pt}
  \setlength{\skip\footins}{3pt plus 1pt minus 1pt}}
\makeatother

 \usepackage{setspace}
\usepackage[colorinlistoftodos,prependcaption,textsize=scriptsize]{todonotes}
\usepackage{todonotes}

\usepackage{blindtext,graphicx}
\usepackage[absolute]{textpos}
\usepackage{datetime}

\definecolor{seagreen}{rgb}{0.18, 0.55, 0.34}
\definecolor{ballblue}{rgb}{0.13, 0.67, 0.8}

\definecolor{darkgreen}{rgb}{0.0, 0.44, 0.34}

\sloppypar

\definecolor{dollarbill}{rgb}{0.52, 0.73, 0.4}

\definecolor{indiagreen}{rgb}{0.07, 0.53, 0.03}

\newcommand\revhid[1]{\todo[linecolor=magenta,backgroundcolor=magenta!15,bordercolor=magenta]{\textcolor{black}{\textbf{#1}}}}

\renewcommand\revhid[1]{}

\newcommand\omf[1]{{\color{black}{#1}}}

\newcommand\marklabel[1]{{\noindent\hyperref[rev:#1]{\markqg{#1}}}}

\hypersetup{
citecolor=blue,
linkcolor=blue,
urlcolor=blue
}

\definecolor{lightblue}{rgb}{0.68, 0.85, 0.9}

\DeclareRobustCommand{\markqg}[1]{{\noindent\textbf{{\sethlcolor{cornflowerblue!40}{\hl{#1}}}}}}

\definecolor{cornflowerblue}{rgb}{0.39, 0.58, 0.93}
\definecolor{brandeisblue}{rgb}{0.0, 0.44, 1.0}

\newcommand\revid[1]{\todo[linecolor=cornflowerblue,backgroundcolor=cornflowerblue!50,bordercolor=cornflowerblue]{\textcolor{black}{\textbf{#1}}}}

\fancypagestyle{firstpage}
{

    \fancyfoot[C]{\thepage}
}

\renewcommand\revid[1]{}

\definecolor{darkspringgreen}{rgb}{0.09, 0.45, 0.27}
\definecolor{lasallegreen}{rgb}{0.03, 0.47, 0.19}
\definecolor{seagreen}{rgb}{0.18, 0.55, 0.20}
\definecolor{coral}{rgb}{1.0, 0.5, 0.31}

\newcommand\icut[1]{} %
\definecolor{gray(x11gray)}{rgb}{0.65, 0.65, 0.65}

\definecolor{deepcarrotorange}{rgb}{0.91, 0.41, 0.17}
\definecolor{forestgreen(web)}{rgb}{0.13, 0.55, 0.13}
\definecolor{burntsienna}{rgb}{0.91, 0.45, 0.32}
\definecolor{cadmiumorange}{rgb}{0.93, 0.53, 0.18}

\usepackage[sort,compress]{cite}
\usepackage{amsmath,amssymb,amsfonts}
\usepackage{algorithmic}
\usepackage{graphicx}
\usepackage{textcomp}
\usepackage{xcolor}
\usepackage{fancyhdr}
\usepackage{hyperref}

\def\BibTeX{{\rm B\kern-.05em{\sc i\kern-.025em b}\kern-.08em
    T\kern-.1667em\lower.7ex\hbox{E}\kern-.125emX}}

\title{\LARGE{\omf{\proposal: Exploiting Dynamic Parallelism\\ in Large Language Model Decoding with a\\Processing-In-Memory-Enabled Computing System}}}

\author{
Yintao He$^{1,2}$ \hspace{0.5em} Haiyu Mao$^{3,4}$ \hspace{0.5em} Christina Giannoula$^{5,6,4}$ \hspace{0.5em} Mohammad Sadrosadati$^4$ \vspace{0em} \\
Juan Gómez-Luna$^7$ \hspace{0.5em} Huawei Li$^{1,2}$ \hspace{0.5em} Xiaowei Li$^{1,2}$ \hspace{0.5em} Ying Wang$^1$ \hspace{0.5em} Onur Mutlu$^4$ \vspace{0em} \\
\normalsize{
$^1$SKLP, Institute of Computing Technology, CAS  \hspace{0.5em} $^2$University of Chinese Academy of Sciences  \hspace{0.5em} $^3$ King's College London}\\
\normalsize{$^4$ETH Zürich \hspace{0.5em} $^5$University of Toronto  \hspace{0.5em} $^6$Vector Institute  \hspace{0.5em} $^7$ NVIDIA
}
}

\begin{document}

\maketitle
\thispagestyle{firstpage}
\begin{abstract}
\begin{abstract}
Large language models (LLMs) are widely used for natural language understanding and text generation. An LLM model relies on a time-consuming step called LLM decoding to generate output tokens. Several prior works focus on improving the performance of LLM decoding using parallelism techniques, such as batching and speculative decoding. 
State-of-the-art LLM decoding has both compute-bound and memory-bound kernels. Some prior works \emph{statically identify} and map these different kernels to a heterogeneous architecture consisting of both processing-in-memory (PIM) units and computation-centric accelerators (e.g., GPUs). We observe that characteristics of LLM decoding kernels (e.g., whether or not a kernel is memory-bound) can change \emph{dynamically} due to parameter changes to meet user and/or system demands, making (1) \emph{static} kernel mapping to PIM units and computation-centric accelerators suboptimal, and (2) one-size-fits-all approach of designing PIM units inefficient due to a large degree of heterogeneity even in memory-bound kernels.

In this paper, we aim to accelerate LLM decoding while considering the dynamically changing characteristics of the kernels involved. We propose PAPI (\textbf{PA}rallel Decoding with \textbf{PI}M), a PIM-enabled heterogeneous architecture that exploits dynamic scheduling of compute-bound or memory-bound kernels to suitable hardware units. PAPI has two key mechanisms: 
(1) \emph{online kernel characterization} to dynamically schedule kernels to the most suitable hardware units at runtime and (2) a PIM-enabled heterogeneous computing system that harmoniously orchestrates both computation-centric processing units (GPU) and hybrid PIM units with different computing capabilities. Our experimental results on three broadly-used LLMs (i.e., LLaMA-65B, GPT-3 66B, and GPT-3 175B) show that PAPI achieves 1.8$\times$ and 11.1$\times$ speedups over a state-of-the-art heterogeneous LLM accelerator (i.e., GPU and PIM) and a state-of-the-art PIM-only LLM accelerator, respectively.
\end{abstract}

\end{abstract}

\section{Introduction}
Large language models (LLMs) have achieved remarkable success across a wide range of applications, excelling not only in natural language processing tasks such as code generation \cite{chen2021evaluating, copilot}, question answering \cite{radford2019language, chowdhery2023palm}, but also image \cite{radford2021, dalle, rombach2022high} and video processing \cite{openai2024sora}. Efficient LLM inference is crucial to unlocking the full potential of these models. 
LLM inference consists of two phases: prefill and decoding~\cite{zhou2024survey}. In the prefill phase, the LLM model processes all input tokens in a request to create hidden states for the decoding phase and generate the first output token. Subsequently, during the conventional decoding phase, the model generates an output token per decoding iteration.

Low execution time of LLM inference is crucial for both user experience in real-time applications and hardware utilization efficiency~\cite{Shen2023Efficient}. The LLM decoding phase dominates the execution time in the LLM inference tasks~\cite{hong2022dfx, zhou2022transpim}. For instance, the serial decoding of the GPT-3 175B model is responsible for 96\% of the overall execution time when the input and output lengths are 32 \cite{choi2023unleashing}. The impact of LLM decoding on overall execution time increases as the output length grows, which is essential for generating more detailed and comprehensive LLM responses~\cite{bai2024longwriter}.

To improve the performance of the decoding phase, prior works employ two main parallelism techniques: batching \cite{yu2022orca, agrawal2023sarathi, patel2024splitwise} and speculative decoding \cite{leviathan2023fast, chen2023accelerating, spector2023accelerating, zhang2023draft}. These techniques enable the generation of multiple tokens, known as \textit{parallel decoding}, in one decoding iteration, to accelerate decoding.
We define the number of decoding tokens that are simultaneously generated as \textit{decoding parallelism}. 
Decoding parallelism directly affects the utilization of memory and computation resources.
As a result, some kernels in decoding become compute-bound, while others become memory-bound.
Recent works explore processing-in-memory (PIM) enabled hybrid designs (i.e., heterogeneous architectures using both PIM units and computation-centric accelerators, like GPUs)~\cite{heo2024neupims, park2024attacc, li2024specpim, seo2024ianus, pan2024instinfer} to accelerate the LLM inference process by mapping compute-bound and memory-bound kernels to computation-centric and memory-centric accelerators. These works \emph{statically} characterize LLM decoding kernels and, based on statically-identified characteristics, schedule different types of kernels to different computation units (e.g., PIM units and computation-centric accelerators like GPUs). 

To study the effectiveness of static scheduling, we profile the kernels used in the decoding phase that employs state-of-the-art parallelism techniques. We observe that some kernels shift from being compute-bound to memory-bound (or vice versa) \emph{dynamically} in response to variations in decoding parallelism. This is because decoding parallelism changes dynamically at runtime. 
There are three main reasons for these dynamic changes in decoding parallelism. 
First, the maximum decoding parallelism in a computing system is limited by the memory requirement of requests, which is dependent on request output lengths and cannot be predicted prior to execution. 
Second, the maximum decoding parallelism is also affected by different user requirements, like quality of service (QoS) \cite{agrawal2024taming}. 
Third, some parallelism techniques \cite{patel2024splitwise, mamou2024accelerating} employ dynamic optimization approaches that adjust the configuration of decoding parallelism (i.e., batch size and speculation length) at runtime to enhance system performance (see Section~\ref{sec:motiv}). 
We conclude that dynamic changes in decoding parallelism cause heterogeneous designs with a \emph{static} scheduling scheme to become suboptimal, as they can mistakenly schedule computation-intensive kernels to PIM units or memory-intensive kernels to computation-centric accelerators (e.g., GPUs).

In this paper, we aim to accelerate parallel decoding by fully leveraging the dynamically changing parallelism properties in LLM inference tasks. To this end, we propose PAPI (\textbf{PA}rallel Decoding with \textbf{PI}M), a PIM-enabled heterogeneous architecture that exploits dynamic scheduling of compute-bound or memory-bound kernels to the most suitable hardware unit for each kernel type.
PAPI's key idea is to enable \emph{online} dynamic task scheduling on a heterogeneous architecture (consisting of GPUs and PIM units) via online identification of kernel properties in LLM decoding. 

PAPI is equipped with three key techniques.
First, we propose \emph{a dynamic parallelism-aware task scheduling framework} to assign kernels to suitable computing platforms at runtime. This approach employs a simple yet effective \emph{kernel bottleneck predictor} with low hardware overhead. 
Second, we design \emph{a heterogeneous architecture with PIM units, GPU, and host CPU} to meet different computing and memory demands of different kernels. 
Third, we design a hybrid PIM architecture that includes two different types of PIM units, i.e., \emph{performance-optimized and memory-capacity-optimized PIM units}, which cater to memory-intensive kernels with different computational demands and memory footprints.

We compare PAPI with the state-of-the-art heterogeneous LLM accelerator composed of AttAcc~\cite{park2024attacc} and 6 NVIDIA A100 GPUs~\cite{choquette2020nvidia} (A100+AttAcc), a heterogeneous architecture composed of Samsung's HBM-PIM \cite{lee2021hardware} and the NVIDIA A100 GPUs (A100+HBM-PIM) and a PIM-only LLM accelerator, AttAcc~\cite{park2024attacc}. 
Our experimental results show that PAPI outperforms A100+AttAcc, A100+HBM-PIM, and AttAcc by 1.8$\times$, 1.9$\times$, and 11.1$\times$, respectively.

This paper makes the following contributions.
\begin{itemize}[leftmargin=4mm,itemsep=0mm,parsep=0mm,topsep=0mm]
\item We observe that parallelism in LLM decoding changes \emph{dynamically}, leading to varying demands in both computation capability and memory bandwidth.
\item We propose PAPI, a PIM-enabled heterogeneous computing system, to meet the different computation and memory bandwidth demands by incorporating both memory-centric PIM units and computation-centric GPU and host CPU.
\item We propose an online parallelism-aware scheduling technique that maps dynamically LLM decoding kernels with different and changing properties to the most appropriate hardware units, including hybrid PIM units.
\item We evaluate PAPI and demonstrate that it provides significant performance and energy benefits over state-of-the-art computing systems for LLM inference.
\end{itemize}

\section{Background}

\subsection{LLM Inference}

\begin{figure}[h!]
\centering
\includegraphics[width=0.9\columnwidth]{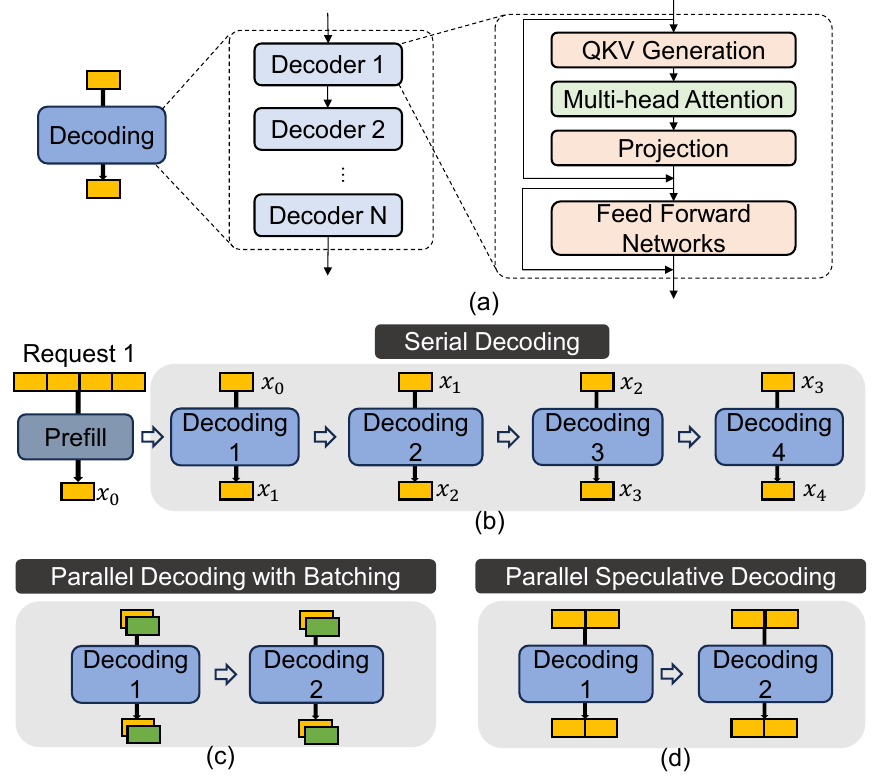}
\caption{(a) LLM structure. (b) LLM inference with serial decoding. (c) Parallel decoding process with batching. (d) Parallel decoding process with speculative decoding.}
\label{fig:parallel_decoding}
\end{figure}

An LLM structure contains several transformer-based decoders, as illustrated in Figure~\ref{fig:parallel_decoding}(a). Each decoder includes four kernels: QKV (Query, Key, and Value) generation, multi-head attention, projection, and feedforward networks \cite{vaswani2017attention}. These kernels can be divided into two types: fully-connected (FC) layers in orange and a multi-head attention layer in green. All kernels consist of general matrix-vector multiplication (GEMV) computations.

Figure~\ref{fig:parallel_decoding}(b) presents an overview of LLM inference, which includes two phases: \emph{prefill} and \emph{decoding}. In the prefill phase, the model processes multiple tokens within the input sequence simultaneously to generate the first output token. The decoding phase has multiple decoding iterations. In each iteration, the model takes the last output token as input to generate a new output token. The output sequence is generated one by one sequentially until the $\text{\textless|eos|\textgreater}$ (end of a sentence) token \cite{vaswani2017attention}. This serialized process is called \emph{serial decoding}. Compared to the prefill phase, the decoding phase typically takes most of the end-to-end inference time \cite{zhang2023draft}. The number of decoding iterations depends on the output token length. With serial decoding, generating $K$ output tokens takes $K$ decoding iterations \cite{leviathan2023fast}.

In LLM inference, the output of QKV generation, i.e., K and V matrices, is stored as these output values are reused multiple times in the multi-head attention kernel during subsequent decoding iterations. In serial decoding, LLM inference requires GPUs/TPUs to load the large weight matrices and KV matrices from off-chip memory to on-chip memory/caches during \emph{each} serial decoding iteration, causing high data movement and performance overheads.

\subsection{Optimization Techniques in LLM Inference}

To overcome performance overheads of conventional serial decoding, researchers have developed two optimization techniques: batching~\cite{vaswani2017attention, yu2022orca, agrawal2023sarathi, patel2024splitwise} and speculative decoding \cite{leviathan2023fast, chen2023accelerating, spector2023accelerating, zhang2023draft}. These methods facilitate the \emph{concurrent} decoding of \emph{multiple} tokens, thereby improving data reuse of weight matrices via generation of multiple tokens, which improves performance.

\subsubsection{Batching}
Batching~\cite{vaswani2017attention, yu2022orca, agrawal2023sarathi, patel2024splitwise} is a parallelism technique to process multiple input sequences concurrently. It allows a single decoding step to generate multiple tokens from different user requests.
This enables request-level parallelism (RLP) during inference, where RLP refers to the number of requests executed in parallel. For example, as shown in Figure~\ref{fig:parallel_decoding}(c), when an LLM processes two input requests simultaneously, RLP is 2. A state-of-the-art batching mechanism is \emph{mixed continuous batching} \cite{agrawal2023sarathi, patel2024splitwise}. In mixed continuous batching, the system dynamically adjusts the number of requests in a batch at runtime based on available memory and computation resources, as well as the number of incoming requests. This technique allows new requests to be added to the current batch without waiting for all requests of the batch to be completed, thereby optimizing resource utilization and improving overall throughput.

\subsubsection{Speculative Decoding} Speculative execution introduces a novel parallel decoding mechanism \cite{leviathan2023fast, chen2023accelerating}, which includes two steps: \emph{serial draft decoding} and \emph{parallel speculative decoding}.
First, a small draft model efficiently predicts the next 2-10 tokens, and these predicted tokens are then verified simultaneously with the large original LLM, allowing for the next tokens to be processed in parallel. Figure~\ref{fig:parallel_decoding}(d) shows this mechanism when two tokens are generated in one decoding iteration of the LLM. Speculative decoding enables token-level parallelism (TLP) in LLM inference.
TLP represents the number of concurrently decoded tokens within a single decoding iteration. For example, two yellow tokens in Figure~\ref{fig:parallel_decoding}(d) for one request are decoded simultaneously, i.e., when TLP = 2.
\section{Motivation}
\label{sec:motiv}

\subsection{Analysis of LLM Inference}

\begin{figure}[b]

\centering
\includegraphics[width=\columnwidth]{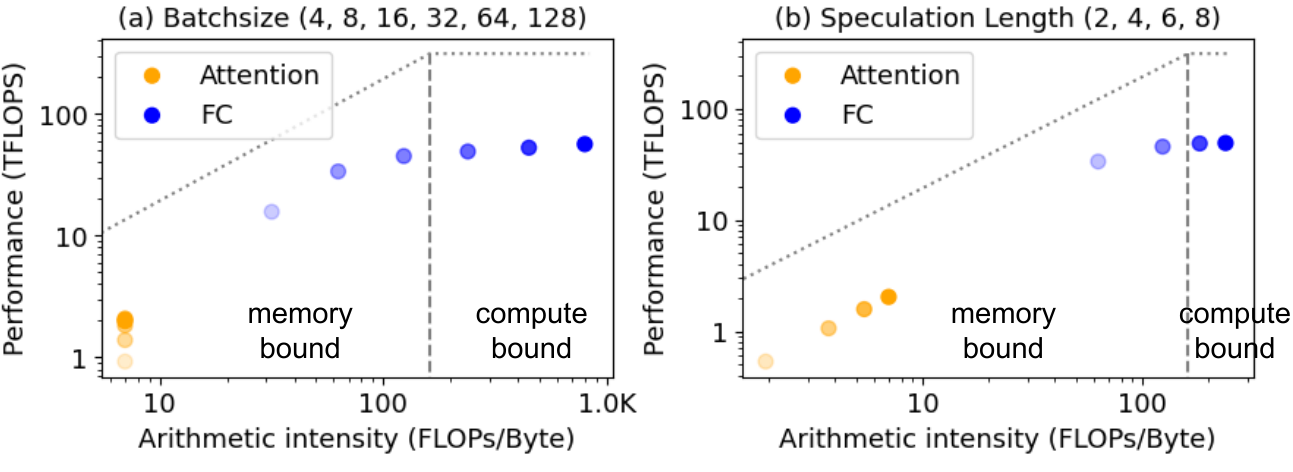}
\caption{Roofline model using OPT-30B with (a) different batch sizes (speculation length = 8) and (b) different speculation lengths (batch size = 32). The darker the color of the dots, the higher the degree of parallelism.}
\label{fig:roofline-model}

\end{figure}

We analyze the computation and memory requirements of LLM inference by evaluating the arithmetic intensity of fully-connected (FC) and multi-head attention kernels. We vary the batch size, when batching is enabled, and speculation length, when speculative decoding is enabled. Figure~\ref{fig:roofline-model}(a) shows the roofline model for the OPT-30B model~\cite{brown2020language} using a high-end NVIDIA A100 GPU~\cite{choquette2020nvidia} with 312 TFLOPS peak computation performance and 1935 GB/s peak memory bandwidth, for FC and attention kernels, as the batch size increases from 4 to 128, with a speculation length of 8. We make two key observations.
First, when the batch size is small, i.e., 4, 8, 16, the decoding phase is memory-bound, i.e., both the attention and FC kernels are bottlenecked by memory bandwidth.
Second, when the batch size is $\geq$ 32, the FC kernel becomes compute-bound, while the attention kernel is memory-bound.
The arithmetic intensity of the FC kernel increases with the batch size. However, the arithmetic intensity of the attention kernel does \emph{not} change when the batch size increases, because there is no data reuse in batching for the attention kernel.

Figure~\ref{fig:roofline-model}(b) shows the roofline analysis of FC and attention kernels when we vary the speculation length from 2 to 8, with a batch size of 32. 
We observe that the arithmetic intensity of the attention and FC kernels increases with the speculation length. With a batch size of 32, the FC kernel becomes compute-bound when the speculation length exceeds 6. In contrast, although the arithmetic intensity of the attention kernel increases with speculation length, the attention kernel remains memory-bound. This is because the arithmetic intensity of the attention kernel increases only slightly with speculation length, and batching has no effect on it, as batching primarily improves weight data reuse, which does not affect the attention kernel.

\subsection{Varying Parallelization Levels in LLM Inference}
\label{sec:3.2}

In real-world LLM tasks, both batch size and speculation length vary significantly at runtime due to changes in user requests and potential adjustments to speculation length to optimize performance~\cite{mamou2024accelerating}. 
We elaborate on why the parallelism in LLM inference dynamically changes during runtime in real-world scenarios.

\noindent\textbf{Initial Request-Level Parallelism (Initial RLP):}
Initial RLP refers to the RLP when batched execution begins. Initial RLP can vary significantly in real-world LLM serving scenarios, causing the batch size to vary greatly. This is due to three major reasons.

\noindent(a) \emph{Service Level Objective (SLO) Limits}: Increasing RLP can enhance throughput but increases inference latency per request~\cite{li2023alpaserve}. Under the online serving scenario, different user latency SLOs dictate varying maximum batch sizes. For example, while a DGX A100 computing system~\cite{DGX} with 1,280 GB memory can support up to 854 requests per batch, a 30 ms SLO requires setting the initial RLP to be as low as 22~\cite{park2024attacc}.

\noindent(b) \emph{Memory Capacity Limits}: Initial RLP is also constrained by the system's memory capacity, particularly for KV cache storage. A computing system with 640 GB HBM can house 282 requests with input and output lengths of 128, but only 18 requests with input and output lengths of~2048~\cite{park2024attacc}. In the latter case, the batch size needs to be smaller, as longer sequences need more memory capacity for KV cache for multi-head attention.

\noindent(c) \emph{Dynamic Batching}: Dynamic batching~\cite{Triton} starts processing a batch once the batch is full or exceeds a time limit. Therefore, when requests are infrequent, an LLM serving system with dynamic batching may start processing with different batch sizes, and thus, different RLP values.

\noindent\textbf{Runtime Request-Level Parallelism (Runtime RLP):}
Runtime RLP refers to the RLP during the execution of a batch of requests. Runtime RLP depends on the batching mechanism used, which may be \emph{static} batching or \emph{mixed continuous} batching \cite{patel2024splitwise}. 

\begin{figure}[H]

\centering
\includegraphics[width=0.7\columnwidth]{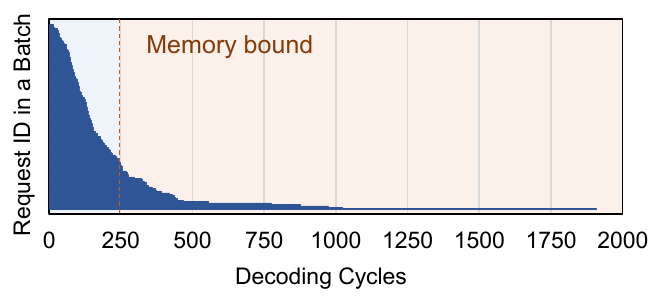}
\caption{Decoding iterations required for each request in a batch, illustrating how the number of remaining parallel requests changes as decoding iterations increase.}
\label{fig:dy}

\end{figure}

Traditional LLM serving systems \cite{Tensorflow, Triton} use static batching with batch-level scheduling. In this approach, no new requests are processed until all requests from the current batch have finished. Since each request has a unique output length, runtime RLP dynamically varies. As shown in Figure~\ref{fig:dy}, runtime RLP dynamically decreases as each request of the current batch finishes (i.e., as more decoding iterations take place)~\cite{oh2024exegpt}.

Mixed continuous batching \cite{agrawal2023sarathi, patel2024splitwise} allows token-level scheduling, where new requests can be added to be processed by the LLM serving system, while the system is executing requests in the current batch. In this case, runtime RLP dynamically changes to keep the hardware resource utilization as high as possible, and it is dependent on when and how many requests are added to each batch.

\noindent\textbf{Token-Level Parallelism (TLP):}
TLP can also be dynamically adjusted at runtime to enhance speculative decoding performance in LLMs~\cite{mamou2024accelerating, su2023synergy}. For instance, a prior work \cite{mamou2024accelerating} introduces a dynamic speculation length optimization technique that adjusts speculation length during each decoding iteration. Additionally, batching and speculative decoding need to be synergistically co-optimized to improve GPU utilization for LLM inference \cite{su2023synergy}: e.g., when the batch size is small, the speculation length can be increased to maximize resource utilization.

We conclude that in real-world LLM serving scenarios, batch size and speculation length substantially vary during runtime.
As a result, the arithmetic intensity of FC and attention kernels dynamically change due to the varying parallelization levels. Therefore, FC and attention kernels can become either compute-bound or memory-bound kernels, when executed in computation-centric systems such as GPUs. We draw two key insights from our analysis.

\noindent\textbf{Key Insight 1.} LLM serving requires a heterogeneous computing architecture with advanced computing units that offer varying computation throughput and memory bandwidth capabilities to satisfy the different arithmetic intensities of kernels. 

\noindent\textbf{Key Insight 2.} LLM serving requires a \emph{dynamic} scheduling approach to map FC kernels to different computing units because FC kernels can switch between being compute-bound or memory-bound during runtime.

\subsection{Limitations of Existing Processing-In-Memory Architectures for LLM Inference}

Computing units, such as GPUs and neural processing units (NPUs), are widely used for LLM serving systems. 
Recent works (e.g., \cite{park2024attacc,seo2024ianus, heo2024neupims,li2024specpim, pan2024instinfer, ortega2024pim, kwon2024lol, gao2024imi, lee2024cost, jeong2024pipepim, zhou2022transpim}) explore the Processing-In-Memory (PIM) computing paradigm (e.g., \cite{aga2017compute, ahn2015scalable, ferreira2021pluto, li2017drisa, seshadri2017ambit, seshadri2015fast, he2021tare, he2020towards, he2019agile}) in LLM inference to alleviate the data movement bottleneck in memory-bound kernels of LLMs, such as the attention kernel.
By integrating processing cores within memory units, PIM provides high memory bandwidth, mitigating data movement bottlenecks in kernels with low arithmetic intensity.

Some prior works~\cite{park2024attacc, seo2024ianus, heo2024neupims, li2024specpim, pan2024instinfer} propose PIM-enabled heterogeneous architectures for LLMs. These architectures include both high-performance computation-centric processors (e.g., GPUs) and PIM devices with very high memory access bandwidth. These works run LLM kernels in computation-centric processors or memory-centric PIM devices and demonstrate significant performance benefits compared to commodity systems, e.g., using only computation-centric accelerators (e.g., GPUs) to run end-to-end LLM inference. However, these prior works still suffer from two major shortcomings.

\noindent\textbf{Shortcoming 1.} \textit{Prior works statically assign FC and attention kernels either to a computation-centric processor (GPU) or a PIM-enabled computing device.} Our analysis shows that \emph{dynamic} assignment of kernels to different computing devices is necessary because LLM kernels have varying arithmetic intensity, e.g., FC kernel can be either compute-bound or memory-bound in GPUs, depending on the speculation length and batch size that are currently used. Specifically, AttAcc \cite{park2024attacc} always offloads all attention kernels to the proposed PIM devices and all FC kernels to a GPU. 
IANUS \cite{seo2024ianus} statically maps all FC kernels to PIM and attention kernels to NPU.
SpecPIM \cite{li2024specpim} proposes an allocation scheme that executes attention and FC kernels at the high-performance processor and PIM devices \textit{concurrently}. However, it is only designed for a static batch size and speculation length.
We conclude that these prior works do \emph{not} sufficiently meet the varying computation and memory demands of real-world LLM serving scenarios. They propose \emph{static} designs, where FC and attention kernels each are always mapped to the same computing hardware; even though kernels exhibit varying computation and memory demands at runtime.

\begin{figure}[H]

\centering
\includegraphics[width=\columnwidth]{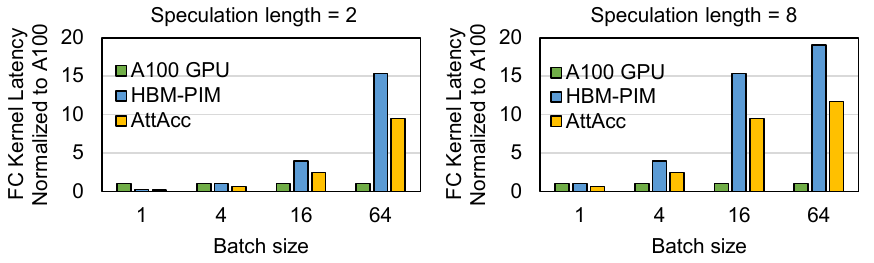}
\caption{The normalized latency of the FC kernel in LLM inference with different parallelization levels (different batch sizes and speculation lengths).}
\label{fig:different-suitable-platforms}

\end{figure}

To quantitatively demonstrate the limitations of prior PIM-based proposals for LLM inference, we evaluate the execution time (latency) of one FC kernel using an NVIDIA A100 GPU \cite{choquette2020nvidia}, Samsung's HBM-PIM architecture \cite{lee2021hardware}, and the state-of-the-art PIM-based work for LLMs, AttAcc \cite{park2024attacc} (Section 7 provides more detail on our evaluation methodology). 
Figure~\ref{fig:different-suitable-platforms} shows the FC kernel latency (normalized to A100 GPU) when we vary the batch size and speculation length.
We observe that in the configurations with low parallelization levels, e.g., having a batch size of 1 and speculation length of 8 or having a batch size of 4 and speculation length of 2, PIM-based architectures, i.e., HBM-PIM and AttAcc, provide better performance than the A100 GPU. In contrast, in the configurations with high parallelization levels, e.g., batch size of 16 or larger, the A100 GPU significantly outperforms the PIM-based architectures, providing much lower execution time.
However, RLP and TLP are \emph{not} known in advance (statically): they dynamically vary and it is hard to predict how they would change. This observation necessitates \emph{dynamic} decisions of which computing hardware to use to execute the FC kernel.

\noindent\textbf{Shortcoming 2.} \textit{Prior works support only one type of PIM-enabled computing device with a certain computation throughput and memory bandwidth capability.} Our analysis shows that although FC and attention kernels can be both memory-bound kernels in GPUs, necessitating PIM-enabled solutions, they have very different arithmetic intensities and different computation and memory bandwidth needs. For example, as demonstrated in Figure~\ref{fig:roofline-model}, with a batch size of 4 and speculation length of 8, the arithmetic intensity of FC is 31.7 FLOPs/Byte, while that of attention is 7.0 FLOPs/Byte. Thus, assuming computing hardware of a certain computation throughput, attention would need around 4.5× higher memory bandwidth than FC. This shows that PIM-enabled computing devices need to provide different computation and memory bandwidth capabilities to efficiently execute the two different types of LLM kernels.
Our work is the first to identify this property of the two types of LLM kernels, while prior works use PIM-enabled devices with a fixed computation and memory bandwidth capability, which make them inefficient at meeting the different and dynamically varying needs of attention and FC kernels.

\begin{figure*}[bp]

\centering
\includegraphics[width=1.82\columnwidth]{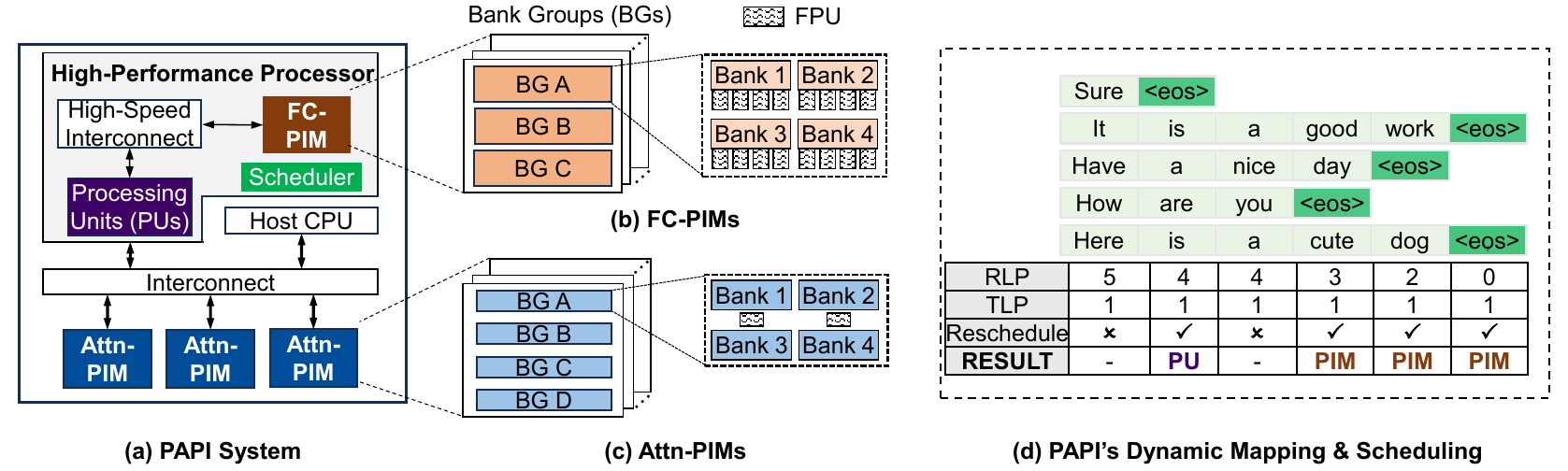}
\caption{Overview of the PAPI computing system, and an example of its dynamic parallelism-aware scheduler.}
\label{fig:heter_arch}

\end{figure*} 

\subsection{Our Goal}
Our goal is to design a versatile computing platform that caters to the varying parallelization levels in real-world LLM inference with different and dynamically changing computation and memory demands. To this end, we propose (1) a heterogeneous architecture that integrates memory-centric PIM units and computation-centric GPU and host CPU, each offering distinct computation throughput and memory bandwidth characteristics, and (2) a parallelism-aware scheduling technique that adapts to runtime variations in parallelization and intelligently and dynamically assigns FC and attention kernels to the most appropriate hardware units in our platform.

\section{PAPI: Overview}

Given that LLM inference exhibits varying parallelization levels during runtime, an intelligent dynamic scheduling policy is necessary to identify the most suitable computing hardware for a given kernel at a given time. 
The \textbf{key challenge} is to design a kernel offloading and allocation scheme that monitors dynamic parallelism online at low cost (in terms of latency and energy consumption) and selects the best-fit computing hardware to fully and efficiently utilize the available hardware resources.

\subsection{PAPI: Key Components}
We propose the PAPI architecture and framework. Figure~\ref{fig:heter_arch} shows the overview of the PAPI framework. PAPI has three key components explained next.

\noindent\textbf{Heterogeneous Architecture.}
We propose a heterogeneous architecture to effectively cater to both compute-bound and memory-bound kernels of LLMs. This architecture includes (1) a host CPU, (2) a high-performance processor with PIM memory units (FC-PIM), and (3) physically separated (i.e., disaggregated) PIM units (Attn-PIM). The high-performance processor includes processing units (hereafter referred to as PUs), e.g., GPU tensor cores~\cite{tensorcore}, PIM memory units (i.e., HBM-based PIM devices), and a hardware scheduler. In our evaluation, we use GPU tensor cores for the PUs, but any other high-performance processor designed for compute-bound kernels (e.g., TPU~\cite{jouppi2017datacenter} or NPU~\cite{chen2014diannao}) could also be used for this design. The host CPU sends instructions to the high-performance processor and the physically separate Attn-PIM devices, which are disaggregated from the high-performance processor.

\noindent\textbf{Hybrid PIM Units.}
We propose two types of PIM units to cater to the different parallelization levels of the FC and attention kernels of LLMs.
FC-PIM units offer relatively high computation capabilities to cater to the FC kernels, while Attn-PIM units provide a larger memory capacity tailored to the attention kernel.
The hybrid PIM units are designed to overcome the limitations of prior existing PIM designs for LLMs (e.g., \cite{lee2021hardware, kwon2022system, park2024attacc}), which typically support a single PIM unit type with fixed computation capabilities.
PAPI separates FC and attention kernels across different PIM devices. 
Since attention kernels are always memory-bound, they are assigned to the Attn-PIM devices.
FC kernels can be either compute- or memory-bound, and thus they can be dynamically allocated by the scheduler to either PUs or FC-PIM units.

\noindent\textbf{Dynamic Parallelism-Aware Scheduling.}
As analyzed in Section~\ref{sec:3.2}, we need to identify whether or not the FC layer is memory-bound and dynamically offload it to the FC-PIM units or the PUs of the high-performance processor. Instead, the attention kernel is always memory-bound, only running on the Attn-PIM units. 
We introduce a hardware scheduler (green block in Figure~\ref{fig:heter_arch}(a)) that monitors runtime parallelization changes and implements dynamic scheduling. When the parallelization level changes, our scheduler executes a low-cost identification step, and offloads the FC kernel to the best-fit computing hardware. 
When the scheduler identifies the FC kernel as memory-bound, it executes FC on the FC-PIM devices. When it identifies FC as compute-bound, it executes FC on the high-performance processor PUs. In the latter case, FC-PIM memory units are used as main memory to keep the weight parameters, which are loaded and processed by the PUs. 
Figure~\ref{fig:heter_arch}(d) illustrates an example of PAPI's dynamic monitoring. Every time the parallelization level of the FC kernel changes, our dynamic monitoring framework is involved, identifying memory-bound or compute-bound kernels, and reallocating them to different units as needed.

\section{PAPI Dynamic Scheduling}

We propose an effective scheduling mechanism to offload FC kernels to PUs or FC-PIM units at runtime with low latency and low energy consumption. In this section, we first explain how the scheduling mechanism determines whether an FC kernel is memory-bound, and then provide the implementation details of the runtime scheduling.

\subsection{Memory-Boundedness Identification of the FC Kernel}

We identify whether or not the FC kernel is memory-bound by estimating its arithmetic intensity. Assume that the weight matrix dimensions of the FC kernel are $(h, h)$ and the input given is $(RLP \times TLP, h)$, where $h$ is the hidden dimension in the LLM structure. The arithmetic intensity of an FC kernel can be calculated as follows:
\begin{equation}
\text{AI} = \frac{\#\text{Flops}}{\#\text{Bytes}} = \frac{RLP \times TLP \times h^2 \times 2}{(2 \times RLP \times TLP \times h + h^2) \times 2}
\end{equation}
In state-of-the-art LLMs, the hidden dimension $h$ is typically large to support their advanced natural language processing tasks \cite{chowdhery2023palm}.
For example, $h = 12288$ in the GPT-3 175B model~\cite{brown2020language}, and the arithmetic intensity can be estimated as follows:
\begin{equation}
\text{AI} \approx RLP \times TLP
\end{equation}
Therefore, we can use $RLP \times TLP$ to estimate the arithmetic intensity of an FC kernel, where $RLP$ and $TLP$ are known at runtime.

\begin{figure}[h]

\includegraphics[width=0.85\columnwidth]{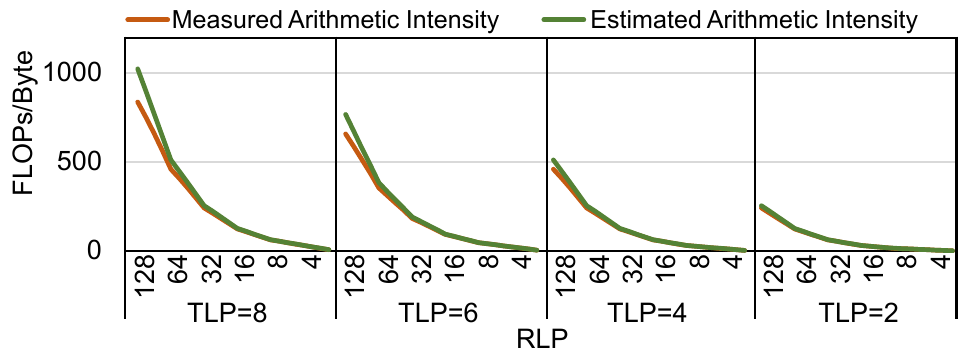}
\caption{Actual measured arithmetic intensity and the estimated arithmetic intensity for FC kernels in the GPT-3 66B model.}
\label{fig:AI-prediction}

\end{figure}

To evaluate the accuracy of our arithmetic intensity estimation, we assess the FC kernel in the GPT-3 66B model using various RLP and TLP configurations. Figure~\ref{fig:AI-prediction} shows the actual obtained arithmetic intensity our estimated values. In most cases, our estimations very closely match the actual arithmetic intensity.
When parallelization level is very large (e.g., RLP=128), the estimated value is slightly larger than the actual arithmetic intensity. In such cases, the actual arithmetic intensity of the FC kernel exceeds the maximum theoretical computation throughput of the PUs of the high-performance processor. Therefore, this small deviation does not impact the offloading decision, correctly identifying the FC kernel as compute-bound and ensuring accurate scheduling.

\subsection{Runtime Scheduling Implementation}
Based on estimated arithmetic intensity, we identify memory-bound or compute-bound FC kernels and dynamically schedule them to the best-fit computing hardware units at runtime.
The scheduling process is executed on the host CPU in two steps: (a) initial scheduling and (b) runtime scheduling.

\subsubsection{Initial Scheduling}
In the initial scheduling step, we decide to offload FC kernels to PUs or FC-PIM units before the LLM serving starts. $RLP$ is set to the batch size, and $TLP$ is set to the system-defined speculation length. We multiply $RLP$ by $TLP$ to estimate the arithmetic intensity and compare it to a memory-boundedness threshold $\alpha$ to make the offloading decision. If the estimated value is larger than $\alpha$, the FC kernel is estimated as compute-bound and offloaded to PUs; otherwise, it is estimated as memory-bound and executed on FC-PIM units. The threshold $\alpha$ is determined through offline iterative evaluation, where we run the FC kernel on both PIM and PU units under varying parallelization levels, using the observed execution times to establish the best $\alpha$ to choose.

\subsubsection{Runtime Scheduling}
In runtime scheduling, we monitor changes in parallelism, predict the current arithmetic intensity, and determine whether or not to reschedule FC kernels to a different computing hardware (i.e., from PUs to FC-PIM units and vice versa). We use a token-level scheduling scheme to track parallelism changes and make real-time decisions based on the estimated arithmetic intensity. The process involves four steps, which we describe next.

First, after each decoding, we gather the output tokens of all requests in the current batch into a single vector.
Second, we count the number of $\text{\textless|eos|\textgreater}$ tokens in this vector to track changes in $RLP$. If the count is greater than zero, it indicates that some requests have been finished, releasing the corresponding PIM resources allocated to Attn-PIM. $TLP$ is typically set initially and does not change frequently at runtime, so we monitor changes in $TLP$ with a direct approach: the $TLP$ value is stored in a dedicated register, and if the system software running on the host CPU modifies $TLP$, the host CPU notifies (sending instructions) the PAPI system to update the register accordingly.
Third, we calculate $RLP \times TLP$ to predict the arithmetic intensity of the next decoding.
Fourth, we compare the estimated value with the memory-bound threshold $\alpha$ to decide whether rescheduling FC kernels from PUs/FC-PIM units to FC-PIM/PUs is needed.
Figure~\ref{fig:heter_arch}(d) shows an example of our proposed dynamic scheduling technique that enables the execution of LLM decoding on the most suitable hardware units of our proposed architecture based on the real-time demands of the workload, significantly optimizing the performance of LLM inference.

\section{PAPI Architecture}

\subsection{FC-PIM Design}
To meet the computation demands of the FC kernel, we need to design a PIM solution with relatively high computation parallelism, while satisfying the necessary power constraints. We modify and use an open-sourced HBM-based PIM simulator \cite{park2024attacc} that is based on Ramulator 2.0~\cite{luo2023ramulator, kim2015ramulator} to evaluate energy consumption and power across different PIM configurations.

\begin{figure}[!t]

\centering
\includegraphics[width=\columnwidth]{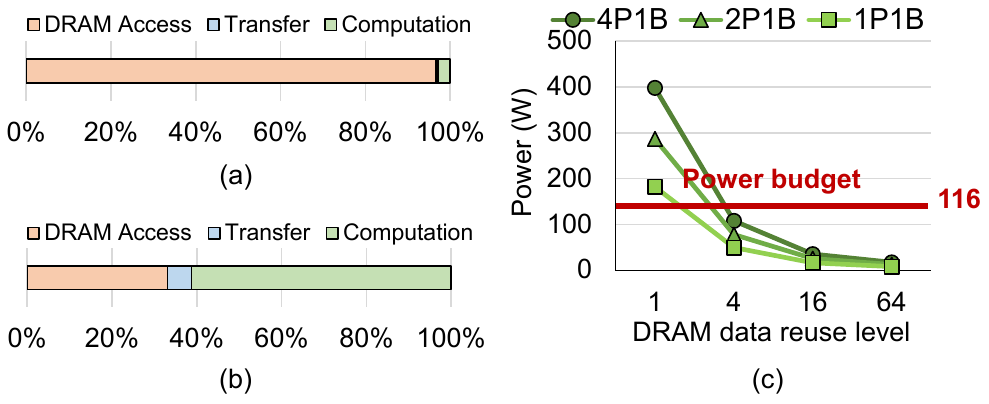}
\caption{(a) Energy breakdown of PIM for executing the FC kernel with no DRAM data reuse. (b) Energy breakdown of PIM for executing the FC kernel when one DRAM access (i.e., an activated DRAM row) is used 64 times for computation (i.e., data reuse level = 64). (c) Power consumption of PIM architecture with different data reuse levels and different numbers of FPUs per bank.}
\label{fig:xPxB} 

\end{figure}

We first examine the energy breakdown in a traditional PIM design (e.g.,~\cite{park2024attacc}) that integrates one processing core per memory bank, referred to as 1P1B.
The energy consumption of PIM execution comes from three parts: $DRAM\; Access$, $Transfer$, and $Computation$. $DRAM\; Access$ includes the energy consumption required to activate and precharge an HBM DRAM row to read the weight data.
$Transfer$ includes the energy consumption of transferring activation data from the buffer die, via the TSV, global controller, and bank group controller, to the processing core.
$Computation$ includes the computation energy in floating point multiplication units (FPUs) of the processing core.
As shown in Figure~\ref{fig:xPxB}(a), most of the energy in PIM execution is consumed by \emph{$DRAM\; Access$}, which accounts for 96.7\% of the total energy consumption\footnote{This energy consumption breakdown is very different from that the HBM-PIM paper~\cite{lee2021hardware} reported. The key difference is that the HBM-PIM paper~\cite{lee2021hardware} reports \emph{only} the energy consumption breakdown of data movement, while we report the energy consumption breakdown of \emph{both} data movement and computation.} larger than the Q vector (activation data).

Based on the above analysis, accessing data from DRAM \emph{once} and reusing it for \emph{multiple} computations can significantly reduce energy consumption. If data can be accessed from DRAM once and used for multiple computations, the total energy consumption of PIM execution can be reduced significantly.
Figure~\ref{fig:xPxB}(b) shows the energy breakdown of PIM when data is fetched once from DRAM and then reused for 64 FC kernel computations.
The energy consumption of $DRAM\;Access$ reduces to 33.1\% of the overall energy consumption. This approach gives us a new opportunity to enhance the parallel computation throughput of near-bank PIM. By lowering the energy cost of DRAM access, we gain additional energy budget for the PIM cores.

As described in Section 2, parallelism techniques (batching and speculative decoding) enable data reuse in LLM decoding, which enables parallel PIM execution by allowing the reduction of the \emph{$DRAM\;Access$} component.
We analyze the power consumption with varying data reuse levels, where a single DRAM access is reused across multiple computations. We explore different PIM configurations, i.e., different numbers of FPUs per DRAM bank. Figure~\ref{fig:xPxB}(c) shows our results. $xPyB$ denotes x FPUs per y banks. The horizontal axis represents the data reuse level, which indicates how many times a single DRAM row is used for FC kernel computations. The vertical axis shows power consumption. We observe that a higher data reuse level leads to a significantly lower power consumption. Specifically, when the data reuse level is $\geq$ 4, the power consumption of 4P1B becomes significantly lower than that without data reuse (i.e., data reuse = 1) and meets the power budget of HBM.\footnote{The power budget of an 8-high, 16GB HBM3 cube is 116 watts~\cite{park2024attacc} following the IDD7 measurement methodology, described in the JEDEC HBM3 specification~\cite{JEDEC}.} Thus, exploiting data reuse enables the use of more FPUs per DRAM bank while staying within power constraints.

Apart from power constraints, the area constraint of a single HBM die is also a significant barrier against highly parallel PIM designs. Thus, we must ensure that the total HBM-PIM die area including the addtional FPUs stays within the maximum allowable area for a single HBM die.
To accommodate additional FPUs within the area-constrained HBM die, we reduce memory capacity, freeing up space for the FPUs.
Assuming each PIM-enabled HBM die has $m$ DRAM banks and each DRAM bank employs $n$ FPUs. The total area of memory and FPUs should satisfy the following condition:
\begin{equation}
    m (n \times A_{FPU} + A_{bank}) \leq A_{Max}
\end{equation}
This equation allows us to calculate $m$ to obtain the maximum capacity achievable in a PIM-enabled HBM die using an nP1B PIM configuration. 

We use the analytical tool CACTI-3DD \cite{chen2012cacti} to estimate area. The area of one HBM bank $A_{bank}$ is 0.83mm²~\footnote{The area of bank includes both the memory array and peripheral circuits.} using a 22nm technology node. The area of one HBM die is constrained to 121 mm² according to prior work \cite{ryu202316}. 
The area of one FPU $A_{FPU}$ is 0.1025 mm² \cite{park2024attacc}.
Thus, the equation for a 4P1B PIM configuration becomes as follows: 
\begin{equation}
    m(0.1025 \times 4 + 0.83) \leq 121
\end{equation}
Therefore, the maximum number of memory banks must be smaller than 97. In our design, we use 96 banks per HBM memory unit, i.e., 3 bank groups (BGs) in the 8-High HBM stack, so as to meet the area constraint of one HBM die with a 4P1B PIM configuration, as shown in Figure~\ref{fig:heter_arch}(b).

\subsection{Attn-PIM Design}

To address the varying arithmetic intensity, computation demands, and memory footprint of FC and attention kernels, while ensuring high hardware resource utilization, we propose dedicated Attn-PIM units, separate from the FC-PIM units (as described in Section 6.1). The Attn-PIM units are disaggregated from the high-performance processor through an interconnect. This disaggregated design of Attn-PIM allows us to tackle the growing memory footprint demands of KV caches of LLMs, as we explain next.

We find that FC kernels of LLMs have larger computation intensity and result in significantly larger latency than the attention kernels, while attention kernels have larger memory footprint demands. Therefore, given a fixed area budget, FC-PIM requires a configuration with higher computation capability, while the attention kernel does not need as much computation capability. To meet these constraints, we allocate FC-PIM devices with high execution parallelism, i.e., 4 FPUs per DRAM bank (as described in Section 6.1). In contrast, we allocate a larger number of Attn-PIM devices, each of which has lower execution parallelism, using 1 FPU for every two banks, as shown in Figures~\ref{fig:heter_arch}(b) and (c), respectively. Using a single FPU for two banks in Attn-PIM devices ensures that power consumption stays within the HBM power constraints. For attention kernels with a speculation length of 1, a single FPU at 666 MHz with 20.8 MB/s per-bank bandwidth (1P1B) matches the arithmetic intensity of the kernel. However, due to the lack of data reuse in this kernel, the power consumption of 1P1B exceeds the power budget, as shown in Figure~\ref{fig:xPxB}(c). Consequently, we adopt the 1P2B configuration for each Attn-PIM device to stay within the power consumption limits.

After configuring the FC-PIM and Attn-PIM hardware designs, we determine how many of each of the two types of PIM devices are required for the entire system to efficiently run LLM inference. We configure the total number of PIM devices in the system considering the capacity requirement of LLM inference.
The memory capacity requirement for the FC kernel is determined solely by the model size and does not change during runtime.
However, the memory capacity required for the attention kernel increases linearly with the sequence length.
To support requests with longer sequence lengths, i.e., requests that produce a larger number of output tokens, we disaggregate the Attn-PIM devices from the high-performance processor, which enables accommodating a large number of Attn-PIM devices that can house large memory footprints (in a flexible manner).

Overall, by separately optimizing the parallel computation and memory capacity capabilities of FC-PIM and Attn-PIM devices and having different numbers of devices in the system for the two types of PIM devices we propose, we can satisfy the higher computation and memory bandwidth demands of FC kernels while also satisfying the higher memory capacity and lower computation demands of attention kernels.

\subsection{System Integration} %
Figure~\ref{fig:heter_arch}(a) shows an overview of the interconnection network between the Attn-PIM devices and the high-performance processor and host CPU, where the high-performance processor consists of FC-PIM devices and processing units. FC-PIM devices require high-speed communication with the processing units due to the large volume of weight parameters transferred.
Therefore, we select high-speed interconnects like NVLink~\cite{NVlink} to connect the FC-PIM devices with the processing units. NVLink provides the required data throughput to ensure that the FC kernels can be executed efficiently without being bottlenecked by data transfer speeds.
In contrast, the attention kernel primarily involves small data transfers, such as byte-level Q vector, so a standard interconnect like PCIe (Peripheral Component Interconnect Express)~\cite{mayhew2003pci} or CXL (Compute Express Link)~\cite{das2024introduction} suffices, depending on the number of devices. PCIe theoretically supports up to 32 devices per bus~\cite{miller2009motivating}, while CXL can scale to 4,096 devices~\cite{das2024introduction}. These conventional links offer adequate bandwidth for attention kernels and are more cost-effective than high-speed ones.

\subsection{Data Partitioning Across PIM Devices}
For the attention kernel, we distribute attention heads across Attn-PIM units, with each head assigned to a separate HBM device. We employ the attention mapping scheme from AttAcc~\cite{park2024attacc} on an HBM device, which ensures efficient data movement and parallelism across the PIM architecture. Specifically, the $K^T$ matrix is partitioned column-wise at the pseudo-channel and bank-group levels, and row-wise at the bank and multiplier level. Conversely, the $V$ matrix is partitioned row-wise at the pseudo-channel and bank-group levels, and column-wise at the bank and multiplier level.

For the FC kernel, the large weight matrix is first divided into smaller 2D blocks, each mapped to an HBM device. At the pseudo-channel, bank-group, and bank levels, these weight blocks are partitioned similarly to the $K^T$ matrix in the attention kernel: column-wise at the pseudo-channel and bank-group levels, and row-wise at the bank level.

\subsection{Practicality and Architectural Scalability}

\textbf{Complementary PIM Units for Diverse Workloads.}
We design different FC-PIM and Attn-PIM devices to address distinct computation and memory access patterns in LLMs while maintaining hardware practicality.
Both FC-PIM and Attn-PIM devices share the same bank-level computation fabric and memory hierarchy. The key difference lies in the number of processing units (PUs) per bank, which is tailored to the specific computation characteristics of LLM tasks. Attn-PIM, optimized for memory-bound operations, uses fewer PUs per bank to handle memory-intensive tasks efficiently, while FC-PIM is designed for more computation-intensive operations like fully connected (FC) layers, with more PUs per bank to enable higher computation throughput.

The design of Attn-PIM has already been demonstrated to be implementable in industry prototypes and products, such as UPMEM~\cite{upmem, upmem2018,gomez2022benchmarking,gomez2023evaluating,rhyner2024pim,hyun2024pathfinding,giannoula2022sparsep,nider2021case,gomez2021benchmarkingcut} and HBM-PIM~\cite{lee2021hardware, kwon202125}. Its integration into our system ensures efficient processing of memory-bound tasks, making it a suitable solution for LLM workloads.

FC-PIM, on the other hand, leverages a higher number of PUs per bank to enhance parallel execution and optimize computation throughput for FC layers, while staying within the power limitations of HBM.

By utilizing a shared HBM-PIM computing substrate, both FC-PIM and Attn-PIM benefit from a unified design that avoids modifications to the DRAM core array. Computation logic is embedded within the peripheral circuits, minimizing area overhead while ensuring compatibility with existing HBM technology. We believe this approach simplifies hardware integration and offers scalability, making PIM technology easier to adapt for large-scale deployment in LLM accelerators.

\noindent\textbf{Deployment of Emerging LLM Models.}
The rapid development of LLMs, particularly Mixture of Experts (MoE) models~\cite{shazeer2017outrageously, lepikhingshard, fedus2022switch}, has introduced new challenges and opportunities for hardware accelerators. MoEs activate only a subset of experts during inference, leveraging sparsity to reduce computation demands. This property is advantageous for hardware accelerators, as it allows for more efficient resource utilization.

FC-PIM is particularly well-suited to exploit the sparsity inherent in MoE architectures. In an MoE model, different experts are activated depending on the input, and the sparsity of these activations presents a significant opportunity to optimize computation. FC-PIM can efficiently execute these sparse operations by storing weight slices from different experts within the same DRAM bank. This allows the system to minimize idle FPUs, which would otherwise remain unused due to the sparsity of MoE models. Moreover, by reducing unnecessary data movement between memory and computation units, FC-PIM helps lower both the energy consumption and the latency associated with MoE inference. These design choices ensure that PAPI can effectively accelerate MoE-based models, making it a viable solution for future LLM architectures.

In summary, the practical implementation of both FC-PIM and Attn-PIM within the PAPI architecture offers a scalable and energy-efficient solution for modern LLM workloads. By leveraging the complementary strengths of these two types of PIM devices and addressing the specific needs of emerging LLM models like MoEs, PAPI is well-positioned to provide high-performance acceleration for a broad range of future LLM applications.

\section{Evaluation}

\subsection{Evaluation Methodology}

\textbf{Comparison Points and Simulation Methodology.} We compare PAPI with three state-of-the-art systems: (a) \emph{A100+AttAcc:} a heterogeneous computing platform with 6 NVIDIA A100 GPUs~\cite{choquette2020nvidia} and AttAcc PIM-based units (one FPU unit per DRAM bank, i.e., 1P1B configuration), which is the state-of-the-art design proposed by prior work \cite{park2024attacc}. All FC kernel computations are executed on GPUs, and attention kernel computations are handled by AttAcc PIM-based units; (b) \emph{A100+HBM-PIM:} an integrated computing platform with 6 NVIDIA A100 GPUs and HBM-PIM devices. HBM-PIM~\cite{lee2021hardware} is a commercial PIM device produced by Samsung, featuring one FPU unit per 2 DRAM banks (i.e., 1P2B configuration); (c) \emph{AttAcc-only:} a PIM-only computing platform with AttAcc PIM-based units~\cite{park2024attacc}, in which all computations of FC and attention kernels are executed on PIM units.
In the PAPI design, the capacity of the FC-PIM devices is 12 GB, while all other HBM devices, including Attn-PIM devices in PAPI, have a capacity of 16 GB. Therefore, one GPU Memory in PAPI is 60 GB rather than 80 GB in the A100 GPU, necessitating six GPUs to accommodate the model parameters of GPT-3 175B (requiring 350 GB memory).
For a fair comparison, each of the computing systems has 90 HBM devices, 30 for storing the weight parameters of FC kernels and 60 for attention kernels. Each GPU contains 5 HBM devices connected via NVLink~\cite{NVlink}, corresponding to the 80GB GPU memory of the A100 GPU.
All HBMs used in the experiments are HBM3~\cite{JEDEC} with 5.2Gbps per pin and running at 333MHz.
We developed a simulator based on Ramulator2~\cite{luo2023ramulator} (new version of Ramulator~\cite{kim2015ramulator}) and AttAcc~\cite{park2024attacc} to evaluate the performance and energy efficiency of the PAPI computing platform, including both GPU and PIM-based components. 

\noindent\textbf{Workloads.} We evaluate three transformer-based LLMs, LLaMA-65B~\cite{touvron2023llama}, GPT-3 66B~\cite{brown2020language}, and GPT-3 175B~\cite{brown2020language}, using the FP16 data type. We use creative-writing and general-qa tasks in the Dolly dataset~\cite{DatabricksBlog2023DollyV2}. The Dolly dataset is an open-source dataset of instruction-following records generated by thousands of Databricks employees in several behavioral categories outlined in InstructGPT~\cite{ouyang2022training}. We use static batching with varying initial request-level parallelism (batch size) across experiments. By evaluating our proposed design on real-world datasets, we can test the performance and energy consumption with various input and output sequence lengths while adapting to dynamic parallelization levels observed at runtime.

\begin{figure*}[bp]

\centering
\includegraphics[width=2.1\columnwidth]{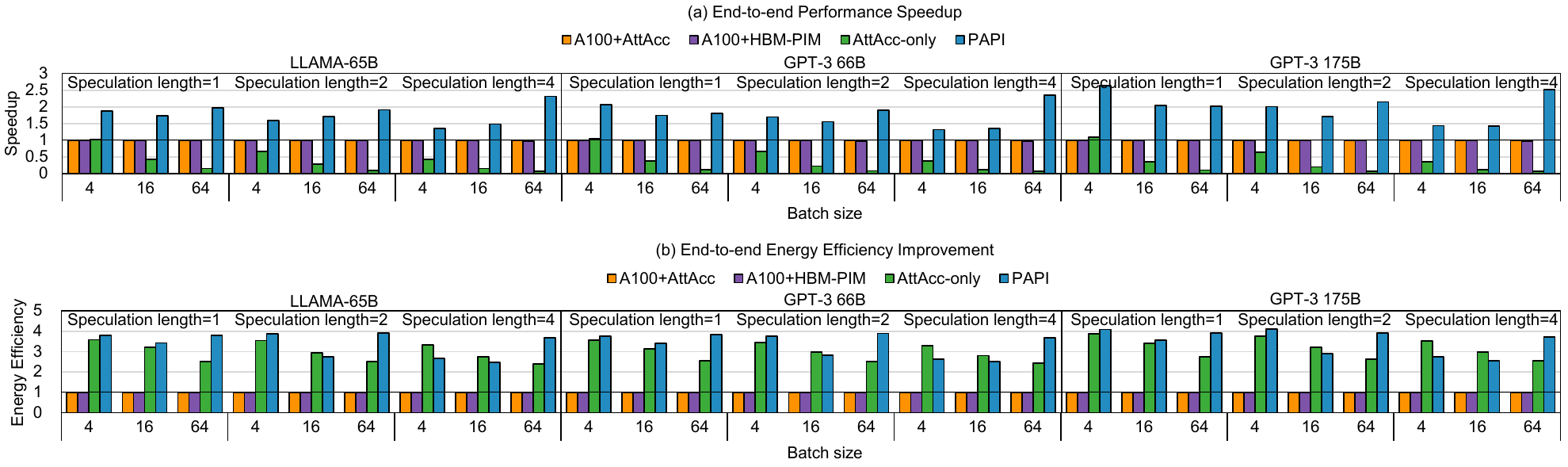}
\caption{End-to-end speedup (top) and energy efficiency (bottom) comparisons of four evaluated designs on the Dolly creative-writing dataset. Values are normalized to A100+AttAcc.}
\label{exp:cw-dataset}

\end{figure*}

\subsection{End-to-End Performance and Energy Efficiency}
\label{sec:7.2}

\noindent\textbf{Performance Speedup.} Figure~\ref{exp:cw-dataset}(a) shows the end-to-end performance of all four evaluated designs using various parallelization levels in the decoding step of each model with batch sizes of 4, 16, or 64 and speculation lengths of 1, 2, or 4. The results are normalized to the A100+AttAcc baseline.

We make three observations. First, PAPI achieves speedups of 1.8×, 1.9×, and 11.1× over A100+AttAcc, A100+HBM-PIM, and AttAcc-only designs, respectively. This is because PAPI schedules tasks between GPU and PIM \emph{dynamically}, offloading each task to the corresponding best-fit computing hardware at a given point in time, and the proposed hybrid PIM architecture can provide varying levels of execution parallelism, catering to the different needs of the FC and attention kernels.
Second, the AttAcc-only scheme performs worse than PAPI and also worse than A100+AttAcc at most parallelization settings. This is due to two reasons. (1) AttAcc-only has limited computation throughput as it employs solely PIM units. Later (Section \ref{sec:7.4}), we compare the performance of PIM solutions with different parallel computation capabilities. 
(2) FC kernels with the parallelism settings used in our experiments are more computation-intensive, making them unable to benefit from a PIM-only solution (which is a better fit for memory-intensive kernels).
Third, A100+AttAcc performs similarly to A100+HBM-PIM because the only difference between them is the execution of the attention kernel on either AttAcc or HBM-PIM. However, the attention kernel's execution time on PIM is relatively small compared to the overall runtime, resulting in a small performance difference.

\begin{figure}[t]
\centering
\includegraphics[width=\columnwidth]{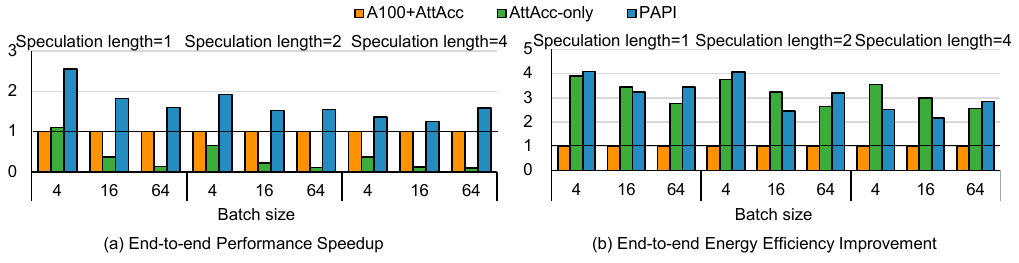}
\caption{End-to-end speedup (a) and energy efficiency (b) comparisons of three evaluated designs on the Dolly general-qa dataset for GPT-3 175B.}
\label{exp:qa-dataset}

\end{figure}

Figure~\ref{exp:qa-dataset}(a) illustrates the end-to-end latency of three designs on the Dolly general-qa dataset. PAPI achieves speedups of 1.7×, 1.7×, and 8.1× over A100+AttAcc, A100+HBM-PIM and AttAcc-only, respectively, which is lower than the speedup for the Dolly creative-writing dataset. This is due to two reasons:
(i) The creative-writing dataset typically has longer output lengths, which makes the decoding phase a larger bottleneck for end-to-end performance, thereby making PAPI acceleration more beneficial.
(ii) Longer output lengths of the creative-writing dataset lead to more significant dynamic changes in parallelization levels, thereby allowing PAPI to further improve performance over prior schemes.

We conclude that PAPI provides significant performance benefits in LLM inference over state-of-the-art PIM-based designs across various real-world configuration settings (speculation length, batch size) and using different real datasets.

\noindent\textbf{Energy Efficiency.} Figures~\ref{exp:cw-dataset}(b) and ~\ref{exp:qa-dataset} (b) present the end-to-end energy efficiency, normalized to A100+AttAcc system, for the creative-writing and general-qa datasets. PAPI improves average energy efficiency by 3.4× and 3.1× for these datasets, respectively, over A100+AttAcc. 
This is because A100+AttAcc executes the FC kernels on energy-hungry A100 GPUs, while PAPI offloads parts of these kernels to FC-PIM devices, thereby consuming less energy by mitigating data movement and exploiting low-power processing cores in memory. 
Compared to AttAcc-only, PAPI provides 1.15× and 1.01× energy efficiency improvement in creative-writing and general-qa datasets, respectively, which is lower than PAPI benefits over A100+AttAcc.
This is because PAPI dynamically schedules the FC kernels on the energy-hungry GPU cores and the energy-efficient PIM cores. While GPU execution consumes more energy than AttAcc-only, PAPI lowers energy consumption on PIM through DRAM data access reuse, resulting in modest savings over AttAcc-only.

We conclude that PAPI improves energy efficiency over state-of-the-art PIM systems across different real configuration settings and datasets.

\subsection{Sensitivity to Parallelization Levels}
We analyze the performance of three evaluated designs across different RLP and TLP values, using the LLaMA-65B model on the creative-writing dataset.

\begin{figure}[h]
\centering
\includegraphics[width=\columnwidth]{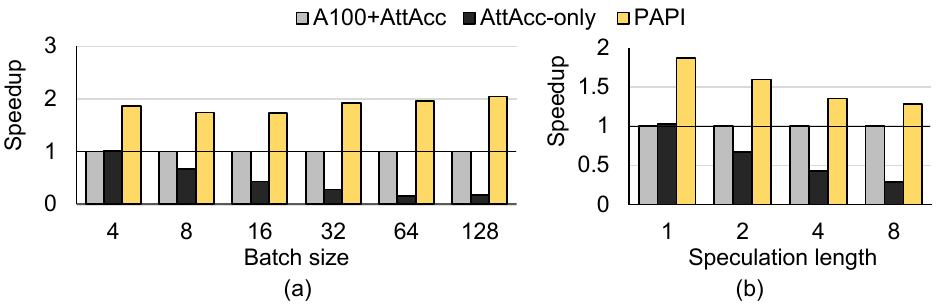}
\caption{End-to-end speedup with (a) different batch sizes (speculation length=1), and (b) different speculation lengths (batch size=4); for LLaMA-65B.}
\label{exp:batch}
\end{figure}

\noindent\textbf{RLP.} Figure~\ref{exp:batch}(a) presents the performance of three designs when we vary the batch size from 4 to 128 to explore the effect of RLP, using a fixed speculation length of 1. When RLP is relatively low, e.g., with a batch size of 4, AttAcc-only provides higher performance than A100+AttAcc. As RLP increases, the execution time of AttAcc-only increases significantly because the PIM devices cannot effectively cater to the large computation needs of the FC kernels, which leads to AttAcc-only providing much worse performance than A100+AttAcc. PAPI achieves the best performance for all RLP settings over state-of-the-art PIM-based systems.

\noindent\textbf{TLP.} Figure~\ref{exp:batch}(b) shows the performance of three designs when we vary the speculation length from 1 to 8 to analyze various TLP levels, using a fixed batch size of 4.
Compared to A100+AttAcc and AttAcc-only, PAPI achieves 1.5× and 3.0× speedup on average. The speedup of PAPI over A100+AttAcc decreases as TLP increases, because PAPI offloads more FC kernels to the GPUs as TLP increases. If TLP becomes large enough, we expect that PAPI would assign all the FC kernels to the GPU and thus the performance of PAPI to converge to that of A100+AttAcc.

\subsection{Performance Analysis of PAPI}
\label{sec:7.4}
\begin{figure}[h]
\centering
\includegraphics[width=0.95\columnwidth]{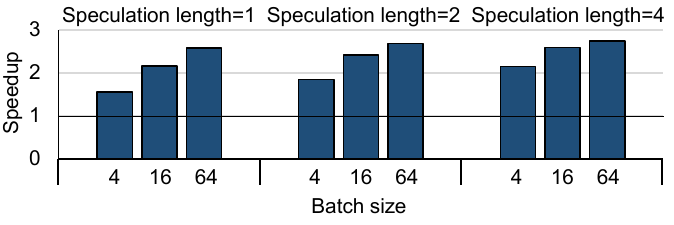}
\caption{Performance speedup of PIM-only PAPI over AttAcc-only in the decoding phase for the Dolly creative-writing dataset.}
\label{exp:pim}
\end{figure}

To analyze the benefits of our proposed hybrid PIM design in PAPI, we compare the performance of two PIM-only architectures: (i) AttAcc-only and (ii) our proposed PIM architecture with only Attn-PIM and FC-PIM devices (but without the A100), using the same number of PIM devices and interconnect settings for fairness.
We only evaluate the decoding phase (since the prefilling phase is compute-bound and is to be executed on the GPU platform).
Figure~\ref{exp:pim} shows the speedup of our PIM-only PAPI design compared to AttAcc-only using the Dolly creative-writing dataset. Our PIM design achieves 2.3× speedup improvement against AttAcc-only on average. We observe that our PIM design has a higher speedup at higher parallelization levels: e.g., when the batch size is 4 and the speculation length is 1, the speedup is 1.6×, while when the batch size is 64 and the speculation length is 4 (higher parallelism) the speedup of PIM-only PAPI increases to 2.7×. This is because, as parallelism increases, FC kernels become more computation-intensive, requiring more computation power. PAPI with more processing units (PUs) can provide the required computation capability much more so than AttAcc-only. Additionally, FC kernels are responsible for most of the execution time, so improving their performance has the largest impact on overall speedup.

\begin{figure}[h]
\centering
\includegraphics[width=\columnwidth]{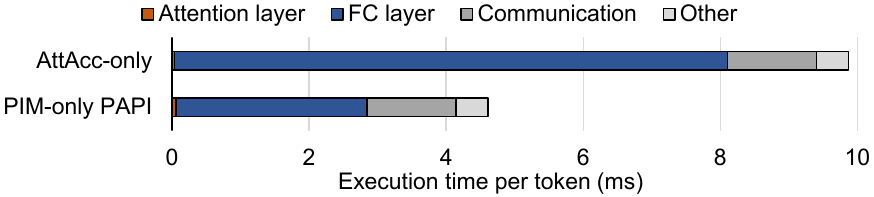}
\caption{Execution time breakdown per token in the decoding phase of LLaMA-65B model inference (batch size=4, speculation length=4) for AttAcc-only versus PIM-only PAPI.}
\label{exp:breakdown}

\end{figure}

Figure~\ref{exp:breakdown} presents the execution time breakdown per token for the AttAcc-only system and for the PIM-only PAPI system with Attn-PIM and FC-PIM devices. We make four key observations.
First, FC kernels dominate the total execution time.
Therefore, it is valuable to enable higher execution parallelism in PIM hardware (as PAPI does with FC-PIM) to effectively cater to the high computation demands of the FC kernels. 
Second, the PIM-only PAPI design provides 2.9× speedup when processing FC kernels.
Third, attention kernels run 1.7× slower on Attn-PIM (1P2B) than AttAcc-only (1P1B) due to our design choice that reduces FPU area overheads.
Fourth, communication takes up 28.2\% of the total execution time in the decoding stage; thus, more advanced network technologies could be developed and integrated into the PAPI architecture to further improve performance. 

\section{Related Work}


To our knowledge, PAPI provides the first architecture and a runtime framework to tackle dynamically varying parallelization levels and hence dynamically varying computation and memory demands of real-world LLM workloads. We comprehensively compare PAPI to two state-of-the-art PIM designs, AttAcc~\cite{park2024attacc} and HBM-PIM~\cite{lee2021hardware}, demonstrating PAPI's significant performance and energy benefits over them (Section \ref{sec:7.2}).


\noindent\textbf{PIM-enabled LLM accelerators.}
The PIM computing paradigm~\cite{mutlu2022modern} addresses the data movement bottleneck between memory and processors by placing computation near or inside memory circuitry.
For transformer-based LLMs, PIM (e.g., \cite{he2020newton, kwon202125, skhynixpim, kwon2022system, park2024lpddr, kal2023aespa, lee2021hardware,jang2024smart, yazdanbakhsh2022sparse, li2024asadi, zhou2022transpim}) provides a promising opportunity to accelerate the memory-bound kernels in the decoding phase.
DRAM-based PIM~\cite{he2020newton, mutlu2024memory}, with its large memory capacity and bandwidth, is particularly well-suited for LLMs. For example, the SK Hynix AiM PIM architecture \cite{kwon2022system} offloads both FC and attention kernels to GDDR6-PIM accelerators, outperforming A100 GPUs in single-batch scenarios. However, the architecture performs poorly when FC kernels are compute-bound, e.g., with larger batch sizes.

Prior works propose heterogenous PIM-enabled computing systems for LLM inference. 
AttAcc \cite{park2024attacc} proposes an HBM-based PIM architecture for attention kernels while running the FC kernels on GPUs to accelerate LLM inference with large batch sizes. Section~\ref{sec:7.2} shows that PAPI outperforms this scheme by designing a more effective PIM-based architecture carefully tailored to the dynamically varying computation and memory needs of FC and attention kernels.
IANUS \cite{seo2024ianus} offloads \emph{all} FC kernels to PIM to efficiently handle non-batched requests.
This would provide low performance in scenarios involving batched requests, which are common in real-world LLM inference.
SpecPIM \cite{li2024specpim} proposes a PIM-enabled system with NPUs and PIM cores, leveraging speculative decoding. It introduces a decoding parallelism-aware scheduling method based on a genetic algorithm and Monte Carlo Tree Search (MCTS). This offline scheduling process involves 50 rounds of the genetic algorithm and 10,000 leaf node searches for MCTS.
While this scheduling method provides performance benefits in cases with a fixed batch size and speculation length, its computational complexity makes it impractical for dynamic execution. In dynamic real-world LLM inference scenarios, especially when decoding parallelism levels vary over time, SpecPIM would need to repeatedly run MTCS scheduling, incurring high-performance costs.

\noindent\textbf{Other LLM accelerators.}
Prior works explore hardware LLM accelerators to improve LLM inference performance.
DFX \cite{hong2022dfx} introduces a multi-FPGA accelerator with high-bandwidth memory (HBM) for end-to-end inference acceleration, and provides an efficient dataflow when the decoding stage is memory-bound. 
However, even when using HBM, such designs still suffer from the memory bottleneck, especially when attention kernels exhibit very low arithmetic intensity \cite{xia2023flashllm}. 
AMX-GPU~\cite{kim2024exploiting} proposes an adaptive LLM model scheduling strategy for CPU-GPU cooperative computing. While this design can adapt to different batch sizes and token lengths, it does not account for runtime changes in parallelism, such as varying concurrency of requests or dynamic changes in computation versus memory bottlenecks.

Recent research utilizes various approximation algorithms, like pruning and quantization, to reduce the amount of data movement (e.g., \cite{wang2023cta, qu2022dota, kao2023flat, ham20203, dong2023heatvit, dass2023vitality, you2023vitcod, ham2021elsa, guo2023olive, lu2021sanger}).
For example, SpAtten \cite{wang2021spatten} introduces token pruning to remove unimportant tokens during inference.
These approximation approaches are suitable for LLM scenarios that can tolerate approximate results. PAPI does not sacrifice quality in LLM serving, while providing significant performance and energy benefits over state-of-the-art systems.

\section{Conclusion}

Real-world LLM services with state-of-the-art parallelism optimization techniques, such as batching and speculation decoding, lead to dynamically-changing parallelization levels. As a result, fully-connected and attention kernels in LLM inference exhibit varying computation and memory demands. To seamlessly adapt to such dynamic demands, we propose PAPI, a computing system that supports three types of computing units with different computation and memory bandwidth capabilities, and a lightweight scheduling framework that offloads fully-connected and attention kernels to the most suitable computing units by monitoring the dynamic parallelization levels in LLM inference at low cost. 
Our evaluation shows that PAPI provides 1.8× and 11.1× performance improvement over state-of-the-art LLM inference systems.
We hope that our work enables further research on leveraging heterogeneous PIM-enabled systems to cater to dynamic real-world execution scenarios in emerging machine learning models such as LLMs.

\section*{Acknowledgements}
We sincerely thank the anonymous reviewers of ASPLOS 2025 for feedback. We thank the SAFARI group members for feedback and the stimulating intellectual environment they provide. This work was supported by the National Natural Science Foundation of China (Grant No. 62090024, 62222411), the Strategic Priority Research Program of the Chinese Academy of Sciences, Grant No. XDB0660100, and the National Key R\&D Program of China, Grant No. 2023YFB4404400. Ying Wang and Huawei Li are the corresponding authors (wangying2009@ict.ac.cn, lihuawei@ict.ac.cn). We acknowledge the generous gifts from our industrial partners, including Google, Huawei, Intel, and Microsoft. This work is supported in part by the ETH Future Computing Laboratory (EFCL), Huawei ZRC Storage Team, Semiconductor Research Corporation, AI Chip Center for Emerging Smart Systems (ACCESS), sponsored by InnoHK funding, Hong Kong SAR, and European Union’s Horizon programme for research and innovation [101047160 - BioPIM].

\bibliographystyle{unsrt}
\bibliography{references}

\end{document}